\documentclass[debug,overfull]{epl}
\usepackage{epsfig}

\title{\bf Pattern formation without heating in an evaporative convection experiment}
\author{Hector Mancini\footnote{e-mail address: hmancini@fisica.unav.es} and Diego Maza}
\institute {Departamento de F\'{\i}sica y Matem\'atica Aplicada. Facultad de Ciencias\\
Universidad de Navarra. 31080 Pamplona, Navarra, Spain.}

\pacs {47.54.+r}  {Pattern selection; pattern formation}
 \pacs {47.20.Dr} {Surface-tension-driven instability}
\pacs  {47.20.Hw} {Morphological instability; phase changes}

\begin{document}

\maketitle

\begin{abstract}
We present an evaporation experiment in a single fluid layer
reproducing conditions of volatile fluids in nature. When latent
heat associated to the evaporation is large enough, the heat flow
through the free surface of the layer generates temperature
gradients that can destabilize the conductive motionless state
giving rise to convective cellular structures without any external
heating. Convective cells can be then observed in the transient
range of evaporation from an initial depth value to a minimum
threshold depth, after which a conductive motionless state appears
until de evaporation finish with a unwetting sequence.
 The sequence of convective patterns obtained here without heating,
is similar to that obtained in B\'enard-Marangoni convection. This
work present the sequence of spatial bifurcations as a function of
the layer depth. The transition between square to hexagonal
pattern, known from non-evaporative experiments, is obtained here
with a similar change in wavelength.

\end{abstract}

\section{Introduction}

Pattern formation in different areas of knowledge had received
great attention in the last decade \cite{cross,alexander}.
Interest in this kind of research arises from the general interest
in nature understanding and also from requirements of industrial
processes like painting, film drying or crystal growth, where
pattern formation knowledge plays a fundamental role. Pattern
formation during evaporation is a common phenomena that can be
frequently observed in nature. Natural convection self-generated
by the evaporation of a thin layer of water, normally left the
brand of its individual convective cells in the bottom clay.

Since the first rigorous work devoted to study pattern formation
in fluids \cite{Benard} the existence of cellular structures was
recognized be linked to surface tension and buoyancy.
    Experimental and theoretical studies where movements
are generated mostly by interfacial forces \cite{block, scriven}
have been increased in the last years
\cite{libroColinet,libroVelarde}. Generally convective movements
originated in surface tension gradients are associated with the
names Marangoni or B\'enard-Marangoni convection (BM). In
evaporative convection there are two main physical mechanisms of
instability relating surface tension gradients, one with a change
in the composition or concentration, so called {\em thermosolutal
convection} and the other with the local dependence of surface
tension with temperature or {\em thermocapillary convection}.

In the first one, experiments are normally performed using
alcohols or other evaporative fluids like polymers in solution
where the proportion between solute and solvent can be changed by
evaporation \cite{DeGennes, fanton}. As an example, Zhang \& Chao
\cite{zhang} presented an experimental work reporting the onset of
patterns considering heating (or even cooling) and evaporation
using this kind of fluids. They used a thin liquid layer of
alcohol (between others liquids) heated from below and the
convective structures have been observed by seeding the fluid with
aluminum powder.

    In pure thermocapillary convection, temperature is involved directly and
movements are now related with local temperature dependence of
surface tension. Hydrodynamics instabilities grows as in the
B\'enard original works, but in our knowledge there are not exist
previous experimental results of pattern formation in evaporation
of pure fluids. Recently Maillard et al. \cite{maillard} presented
a microscopic  evidence of unusual patterns (micron sized objects
like rings and  hexagonal arrays) that they consider patterns of
B\'enard-Marangoni convection driven by surface tension gradients.
Even if the work is not performed with an evaporative fluid, it
certify that interest is arising from material science in the
production and control of well-ordered arrays of this kind of
cells at micron scale.

\begin{figure}

\onefigure [scale=0.8] {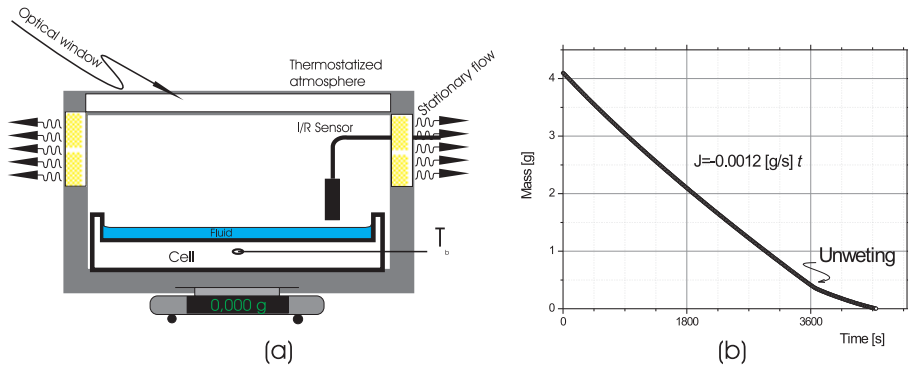}

\caption{(a) Scheme of the experiment: a cylindrical cell is
mounted in a closed environment where a very small flow refresh
air and vapor pressure.(b) The pressure unbalance drive an almost
linear evaporation rate until the drying process begins.}

\end{figure}

Regarding properties of evaporation there is a family of
experimental and theoretical studies aiming to determine constants
like the Sherwood number (the adimensional rate of evaporation)
under different conditions, \cite{sparrow, saylor} or the
temperature profile near the evaporating surface \cite{fang,
bedeaux}. Normally, pattern formation is not an specific object of
these works and must be said that the simultaneous measurement of
all parameters involved is normally a complex task.

    In all the experiments above mentioned, patterns are composed
 mostly by irregular cells. Normally aspect ratio is large (we call aspect
ratio $\Gamma$, the proportion between the horizontal characteristic
dimension and depth). In the former experiments, to our knowledge,
no attempts has been made to compare evaporative experiments with
theory in the frame of pattern formation. The present work is the
first experiment devoted to check the convective pattern sequence
appearing in an evaporation layer without heating and to compare
it with non evaporative convection.

\section{The experiment}

It is well known from thermodynamics that equilibrium in a fluid
with its vapor phase is bidimensional. The equilibrium states are
all on a line in the plane defined by pressure {\em p}, and
temperature {\em T}. In a closed environment with a layer of
volatile fluid, the fluid evaporates until the vapor phase reaches
the vapor pressure corresponding to the fluid temperature. When
the equilibrium pressure is reached evaporation stops. If a part
of the atmosphere composed by air plus vapor is removed (i.e.
blowing slightly), the fluid tends to evaporate continuously until
recover its equilibrium value. If the volume of air is replaced at
the same rate that the mixture of vapor plus air is removed, a
constant evaporation rate can be reached. Under this condition
evaporation follows until all the liquid layer disappear. This a
common situation in nature when the wind removes the vapor phase
in equilibrium with a fluid and makes to evaporate completely a
volatile layer.

In the experiment here presented we reproduce this conditions
introducing all the set-up in a closed box and evacuating a small
part of the total volume of the inner box atmosphere. A scheme of
the system can be seen in Fig. 1.a.

 The fluid used was hexametildisiloxane ($C_{6}H_{18}OSi_{2}$)
and it was choosed because it is a pure volatile fluid at
atmospheric pressure (Prandtl number $= 14.5$). Properties of this
and other silicon oils can be obtained from different handbooks
\cite{handbooks, nist}. It was placed in a cylindrical container
on an electronic analytical balance in order to measure in real
time the mass evaporated (with a precision of $0.001\,g$). The
atmosphere at the inner of the box, composed by the vapor pressure
of the fluid plus air, was kept at a constant pressure by
refilling with a laminar flow of new air at the same temperature
and at the same rate of evacuation. This stationary state,
slightly out of equilibrium, generates a constant rate of
evaporation. The evaporation rate \textbf{J} is defined by the
rate at which the vapor pressure is removed from the gas phase and
by the temperature of the fluid phase.

There is no external heating or cooling in this experiment. The
latent heat is the responsible for convection. The pool is left to
reach its own thermal equilibrium with the environment in a time
that depends on its thermal conductivity. Thermal exchanges are
restricted to different parts of the system at the inner of the
box. The box with all the system is placed in a conditioned air
room in order to keep constant also the external temperature. The
latent heat extracted from the fluid is subsequently recovered
from the surroundings of the layer. Two extreme conditions of
conductibility had been used in the container of the fluid to
check their influence. One was a good conductor aluminum cell and
other with equal dimensions and geometry but constructed in a
thermal isolator material. We do not found significative
differences in the patterns obtained further the different
recovery times.

    In the experiment, patterns appears spontaneously if the
evaporation rate and the fluid depth are adequate. To obtain
ordered patterns the evaporation rate \textbf{J} must be
relatively low. In a typical sequence \textbf{J} was controlled
between $0.0010\, g/s < \mathbf{J} < 0.0015\, g/s$. Depth of the
fluids are obtained from this value, the volume of the cylindrical
container (Area $=80.12 cm^2$), and the fluid density ($0.760\,
g/cc$). A typical evaporation rate obtained in the experiment can
be seen in Fig 1.b. It is interesting to note that only when the
drying process of the layer begins there is a sudden change in the
evaporation rate.

Before the unwetting (or drying process) and without refilling the
cell with new fluid, the depth of the layer changes linearly from
a fixed and arbitrary initial value to zero, independently of what
kind of patterns appear.

\section{Pattern dynamics}
    Patterns are observed by a usual shadowgraph techniques
described in other former experiments of the authors
\cite{hector}. The transient sequence of patterns shown in Fig.2,
have been obtained when a linear change in depth with time is
prepared in the experiment. We used an image processing system to
captures images that then are processed and stored together with
the corresponding outputs of the data acquisition system. Images
and data files of temperatures, depth and evaporation rate
obtained at the same time are then used to obtain the results here
presented. The error in time synchronism of all the system in
negligible. To control the results, each experiment has been
performed more than 20 times. It was verified that if the
evaporated mass is refilled to the pool keeping depth constant, a
really stationary pattern can be obtained.  As shown in Fig. 1.b,
the change in depth at a fixed evaporation rate fits an almost
linear function. The second order coefficient (deviation from
linearity) is obtained reproducibly and is two orders of magnitude
lower than the linear coefficient.

\begin{figure}

\onefigure [scale=0.7] {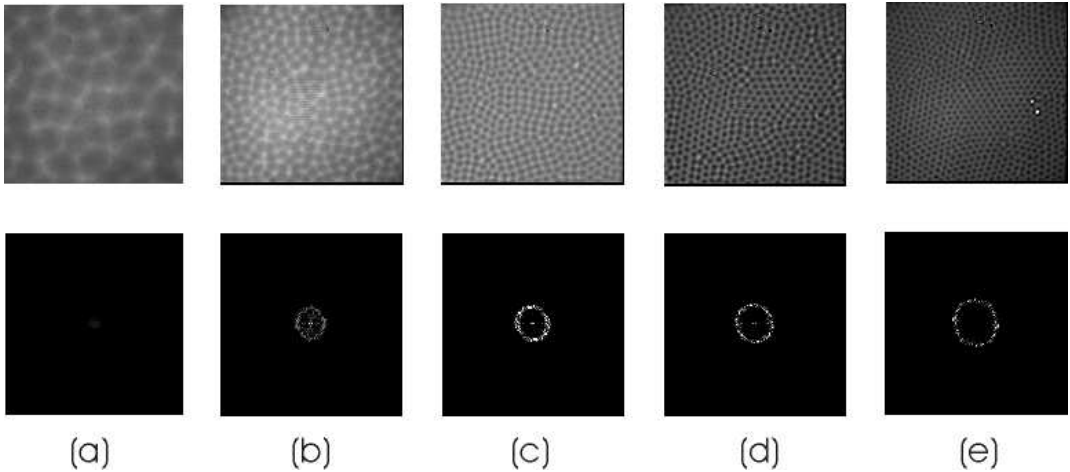}

\caption{Sequence of patterns obtained as a function of decreasing
depth and its FFT spectrum.(a) $d=0.8\,mm$. (b)$d=0.7\,mm$.
(c)$d=0.6\,mm$. (d)$d=0.5\,mm$. (e)$d=0.4\,mm$.}

\end{figure}

The sequence begins when the cell is filled with a fixed and
arbitrary initial depth of fluid (normally we used $d_0 = 2\,
mm$). Considering the diameter of the cell it means an initial
aspect ratio of $\Gamma \approx 50$. Initially, convective
movements are in turbulent phase. Movements are mostly formed by
thermals which born at the bottom of the cell and appear randomly
distributed in space and time. When the layer depth goes under a
certain value, typically $d= 0,8\, mm$, a pattern formed by a few
irregular and large cells fills the pool (Fig. 2.a). Lowering the
layer depth, the size of the cells is lower and consequently the
number of cells increase (Fig. 2.b). The planform changes
continuously with depth to a pattern composed by tetragonal cells
(Fig 2.c).

\begin{figure}

\onefigure [scale=0.7] {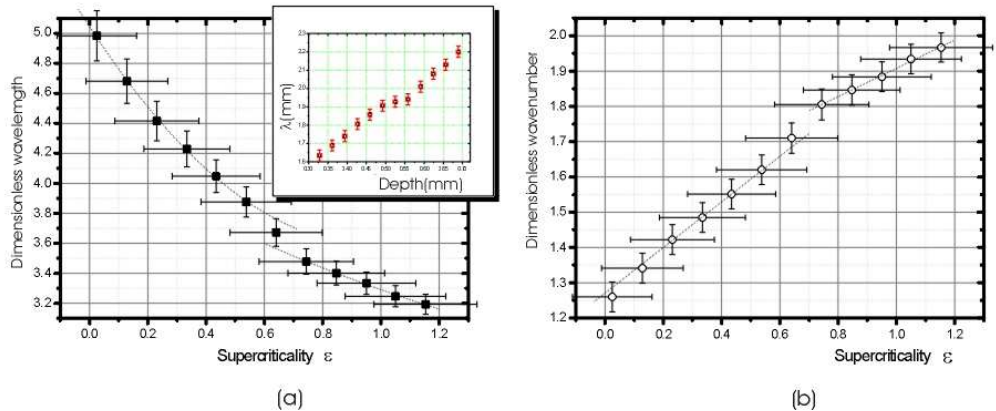}

\caption{(a)Mean dimesionless wavelength as a function of the
distance to the critical depth $d_{c}$. The inset show the
dimensional wavelength as a function of the liquid depth. Note the
existence of a transition zone corresponding to the square to
hexagon transition. (b) Dimensional wavenumber as a function of
the supercriticality. The fits are just guide for the eye.}

\end{figure}

If depth goes down to $d = 0.5\,mm$ approximately, suddenly the
pattern composed by domains of tetragonal cells of Fig. 2.c
changes to other of hexagonal cells like in the Fig.2.d, that
exist until a critical depth $d_c$ is reached. When the minimum
depth is reached all the pattern disappear until that "drying
process" begins. Drying means destruction of the layer and begins
when a long wavelength instability appears giving place to other
different stage in the experiment. The FFT sequence displayed in
Fig. 2, show the existence of a well defined wavenumber but
without a preferred direction in the phase plane.

Fig. 3.a display a typical data file of the mean wavelength
against supercriticality in a run. We have defined here the
supercriticality $\epsilon=\frac{d-d_{c}}{d_{c}}$ as the parameter
distance from the critical depth $d_{c}$ in order to compare with
the results obtained in a non-evaporative convection with heating
\cite{libroColinet}.

Experiments on square to hexagon transition ($SHT$) has been only
recently reported  \cite{michael,vanhook}. This transition shows
strong hysteresis effects. The wavelength at the transition of
hexagons to squares (when $\epsilon$ is increased) is $10\%$ lower
than the wavelength when the supercriticality is diminished. In
our experiment, such hysteresis effect can not be measured. We
observe a "jump" in the dimensionless wavelength when the $SHT$
take place (Fig 3.a). Note from the inset in the Fig 3.a, that the
medium wavelength of the cellular pattern remains almost constant
for depth between $0.49\,mm< d< 0.56\, mm$ .This fact implies
approximately a jump of $5\%$ in the wavelength normalized by the
depth of the fluid.

Within our optical test resolution, we determine almost the same
critical depth $d_{c}=0.3\, mm$, for all the evaporation rate
studied. The dimensionless wavenumber $k_{c}=1.25 $ corresponding
to the onset (Fig 3.b), is in disagreement with the predicted
value for the linear theory \cite{carlos,schatz} which is
$k_{c}=1.97$. This disagreement can be related with the fact of
the convective pattern at threshold not is stationary but have a
characteristic life-time related with the evaporation rate {\bf J}
and the horizontal diffusion time. A more advanced experimental
setup, where the liquid depth remains almost constant during a
horizontal diffusion time, is been implement in order to study the
critical wavenumber at the onset. These results will be reported
elsewhere.

Other outputs of the experiment are the local temperatures against
time (or depth). To measure local temperatures at different
depths, sub millimeter thermocouple are mounted laterally in a
small hole on a plexiglass ring to avoid a pattern perturbation.
Plexiglass is a thermal isolator having a similar conductivity to
the silicon oil, so temperature recovering from the environment is
obtained principally from the aluminum disk placed below. An
infrared sensor (IR) is used to measure the temperature at the
free surface.  From the thermocouple putting inside the cell, we
determine temperature dynamics. Two regimes can be identified:
{\em (a)} when the fluid is in turbulent regimen, the temperature
difference between fluid and atmosphere rise to a limit value
between $-0.8\, ^{o} C$ to $-1.0\, ^{o} C$ and  {\em (b)} when the
stationary pattern stabilizes, this difference decrease to $-0.4\,
^{o}C$ and remains almost constant to the unweting process.Finally
the temperature  goes up to the initial equilibrium value.

\section{Discussion}

In this experiment appear a typical sequence of ordered patterns
similar to a B\'enard-Marangoni convection heated from below. As
the sequence appear here without any external heating, only very
few of the previously existing models describing evaporation
patterns are useful. The typical cellular patterns obtained can be
observed and measured reproducibly. As during the evaporation the
fluid layer depth goes down, the control parameter is consequently
lower and the sequence obtained is inverted in order respect to
normal BM convection.

The first effect produced by evaporation in our experiment is to
create the vertical temperature gradient by latent heat. So we do
not need any heating or cooling flow. The second one is to
increase the thermal conductivity in the evaporating surface. It
means to increase the cooling in the cold surface points and the
heating in the hot one's. This in turn increases the effective
Biot number in the models. It can be verified experimentally by
observing that the system become strongly turbulent when the
evaporating rate is increased too much. To have ordered patterns
the flow of mass must be controlled.

Normally the theoretical models for this kind of experiments
consider an externally imposed heating or cooling flow (from below
or from above). Very recently appeared a theoretical work of Merkt
\& Bestehorn \cite{merkt} where this sequence is obtained without
external driving. They constructed two theoretical models, one is
two layer (fluid and gas) approximation where they perform a
linear stability analysis. The other is a one layer approximation
with a large effective Biot number.

With the second one they found that thresholds obtained in
non-evaporating oils with a fixed Prandtl number fluid (Pr= 10)
are significatively lowered with increasing the Biot number. The
pattern morphology is reproduced and the sequence of bifurcations
obtained in our experiment is then reproduced numerically (even if
there are some differences in the numerical values). The high
effective Biot number hypothesis is very acceptable for us and
also is reinforced because we use an small air flow to carry
sligthly out of equilibrium the system. This must further increase
the heat exchange, equivalent at to have a better heat conduction
in the top surface (an even higher Biot number).

\section{Conclusions}
    In summary, we presented here the first experimental report of
ordered spatial bifurcations produced only by evaporation and a
time solved information of the relevant variables. The sequence of
pattern described is the same as in non-evaporative convection for
a control parameter increasing its value. The main features of the
experiment can be explained by a recent theory \cite{merkt}. We
demonstrate that patterns exist in a well defined range of depths
and also that the transition between squares to hexagons appear as
clearly as in non-evaporative convection, with a change in the
dimensionless wavenumber similar to the value reported for non
evaporative convection. The tetragonal structure appears when
convection is more important to the heat transport (higher depth)
and this result confirm also that tetragonal cells seem to be more
efficient than hexagonal in heat transport,as was assumed in
\cite{michael}.

\acknowledgments

    Authors thanks to M. Bestehorn for share with us their theoretical
results and for accept to exchange it with our experimental
results before publication. Also we are indebted to Iker Zurigel
for his participation in the first stages of this experiment and
Emmanuel Mancini for the cooperation in the construction of the
experimental set-up and during the development of  measurements.
This work was partially supported by the European Community
network ECC TRN HPRN-CT-2000-00158 and by MCyT proyect
BFM2002-02011, Spain.


\begin{thebibliography}{0}

\bibitem{cross}
 \Name{ Cross M. C. \and Hohenberg P. C.}
 \Review {Rev. Mod. Phys.}\Vol{65}\Year {1993}\Page {851}

\bibitem{alexander}
\Name {Rabinovich M., Ezersky A. \and Weidman P.}
 \Book{The Dynamics of Patterns}
 \Publ{World Scientifir, Singapore}
  \Year{2000}

\bibitem{Benard}
\Name{B\'enard H.} {\em Rev. G\'e. Sci. Pures Appl.}  {\bf 11}
\Year {1900} \Page {1261}

\bibitem{block}
\Name{Block M.} {\bf 178} \Year {1956} \Page {650}

\bibitem{scriven}
\Name {Scriven L.E. \and Sterling C.V.}  {\em Nature} {\bf 187}
 \Year{1960} \Page{186}

 \bibitem{libroColinet}
\Name { Colinet P., Legros J.C. \and Velarde M.G.}
 \Book{Nonlinear Dynamics of Surface-Tension-Driven Instabilities}
 \Publ{Wiley, Berlin}
  \Year{2001}

\bibitem{libroVelarde}
\Name {Nepomnyashchy A.,Velarde M. G. \and Colinet P.}
 \Book{Interfacial Phenomena and Convection}
 \Publ{Chapman \and Hall-CRC, EE UU}
  \Year{2002}

\bibitem{DeGennes}
\Name{De Gennes P-G.} arXIV: cond-mat/0111167 \Year{2001}

\bibitem{fanton}
\Name{Fanton X. \and Cazabat A. M. }
 {\em Langmuir} {\bf 14} \Year{1998} \Page{2554}

\bibitem{zhang}
\Name{Zhang N. \and Chao D. F.} {\em Int. Comm. Heat Mass
Transfer}{\bf 26} \Year{1999} \Page{1069}

\bibitem{maillard}
\Name{Maillard M., Motte L., Ngo A. T. \and Pileni M. P.} {\em J.
Phys. Chem. B} {\bf 104},\Year{2000} \Page{11871}

\bibitem{sparrow}
\Name{Sparrow E. M. \and Nunez G. A.} {\em Int. J. Heath and Mass
Transfer} {\bf 31} \Year{1988} \Page{1345}

\bibitem{saylor}
\Name {Saylor J. R. , Smith G. B. \and  Flack K.A.}  Phys. Fluids
{\bf 13} \Year{2001} \Page{428}

\bibitem{fang}
\Name {Fang G. \and Ward C. A.} {\em Phys. Rev.}{\bf E 59}
\Year{1999} \Page{417}. \Name {Fang G. \and Ward C. A.} {\em Phys.
Rev.}{\bf E 59} \Year{1999} \Page{429}. \Name {Fang G. \and Ward
C. A.} {\em Phys. Rev.}{\bf E 59} \Year{1999} \Page{441}

\bibitem {bedeaux}
\Name{Bedeaux D. \and Kjelstrup S.} {\em Physica}{\bf A 270},
\Year{1999}\Page{413}

 \bibitem {handbooks}
  \Book{Polymer Data Handbook. J. E. Marx, Edit. }
   \Publ{Oxford Univ. Press New York}
    \Year{1999}


\bibitem{nist}
  \Book{NIST Chemistry WebBook }
   \Publ{http://webbook.nist.gov}
    \Year{2003}

\bibitem{hector}
\Name{Ondar\c{c}uhu T., Millan-Rodriguez J., Mancini H. L.,
Garcimart\'in A. \& P\'erez Garc\'ia C.} {\em Phys. Rev. E}{\bf
48}\Year{1994}\Page{1121}

\bibitem{michael}
\Name {Eckert K., Bestehorn M. \and Thess A.} {\em J. Fluid Mech.}
{\bf 356} \Year{1998} \Page{155}.

\bibitem{vanhook}
\Name {VanHook S., J.Schatz M. F., Swift J.B., McCormick, W. D.
\and Swinney H.L.} {\em Phys. Rev. Lett.} {\bf 75} \Year{1995}
\Page{4397}.

 \bibitem{carlos}
 \Name {Perez Garcia C., Echebarria B., Bestehorn M.}
 {\em Phys. Rev}. E {\bf 57} \Year{1998} \Page{475}.

\bibitem{schatz}
\Name {Schatz M. F., VanHook S. J., McCormick, W. D., Swift J.B.
\and Swinney H.L.} {\em Phys. Fluids} {\bf 11} \Year{1999}
\Page{2577}.

\bibitem{merkt}
 \Name{Merkt D. \and Bestehorn M.}{\em Physica {\bf D185}}
  \Year{2003}\Page{196}







\end{thebibliography}
\end{document}